\documentclass[11pt]{article}

\input epsf.sty

\usepackage{graphicx,amsmath,amssymb}
\def\lromn#1{\uppercase\expandafter{\romannumeral#1}}

\begin{document}

\begin{center}
\begin{Large}
\textbf{
Primordial neutron star; a new candidate of dark matter
}
\end{Large}

\vspace{1cm}

M. Yoshimura

Research Institute for Interdisciplinary Science,
Okayama University \\
Tsushima-naka 3-1-1 Kita-ku Okayama
700-8530 Japan

\vspace{7cm}

{\bf ABSTRACT}

\end{center}

\vspace{1cm}

Z-boson exchange interaction induces attractive force between
left-handed neutrino and neutron.
The Ginzburg-Landau mean field calculation
and the Bogoliubov transformation
suggest that this attractive force leads to neutrino-neutron pair condensate and 
super-fluidity.
When the result of super-fluid formation is applied to the early universe,
horizon scale pair condensate may become  a component of dark energy. 
A further accretion of other fermions from thermal cosmic medium gives a seed of
primordial neutron stars made of 
proton, neutron, electron, and neutrino in beta-equilibrium.
Primordial neutron stars may provide a mechanism of giving a part or the whole of
the dark matter in the present universe, if a properly chosen small fraction
of cosmic thermal particles condenses 
to neutrino-neutron super-fluid and primordial neutron star not
to over-close the universe.  
The proposal can be verified in principle
 by measuring neutrino burst at primordial neutron
star formation and by detecting  super-fluid relic neutrinos 
in atomic experiments at laboratories.

\vspace{4cm}

Keywords
\hspace{0.5cm} 
Neutrino interaction,
Super-fluid condensate,
Primordial neutron star,
Dark matter

\newpage

\section
 {\bf Introduction}

Neutrinos rarely interact, but their feeble interaction with matter
has been very useful to clarify the point-like structure of nucleons, namely
existence of quarks, using large scale detectors in accelerator experiments.
More dramatically, the standard cosmology predicts that
neutrinos frequently interact in dense thermal medium
of early universe, and  their properties are well described by the
thermalized momentum distribution of zero chemical potential
when the universe is hot enough at temperatures above 1 MeV.
This has been a basis of nucleo-synthesis calculation since the pioneering 
work of C. Hayashi \cite{hayashi}.
Another place one expects  neutrino equilibrium is
deep inside neutron stars after gravitational collapse of iron-rich stars.

In the present work we  explore a novel possibility of
super-fluid phase transition  due to interaction between neutrino and
other fermions.
The description of ideal gas distribution is found to
require a  revision from the simple description, and we may study implications
of super-fluid neutrino-neutron pair condensate.

Our basic theoretical framework is the
well established standard theory of
particle physics, however modified by finite neutrino masses.
How neutrino masses, either of Majorana or Dirac type,
are generated is irrelevant to the present work.
Within this framework we work out neutrino interaction
by Z-boson exchange  and find that it is given by an attractive force
when the interacting partner is neutron.
It is further shown that Z-boson plays a similar role to phonon
in super-conductive metal \cite{superconductivity}, and
left-handed spin-singlet pair $\nu_L n_L$ condensates like
Cooper pair.
Theoretical tools we use are the Ginzburg-Landau mean field theory
and the technique of Bogoliubov transformation, both of which
are standard methods in the theory of superconductivity.

Accretion of ambient thermal fermions on super-fluid $\nu_L n_L$ condensates
in the early universe
suggests that they give rise to a seed of dark matter, 
primordial  primordial neutron stars made of neutrino, electron, proton
and neutron in beta equilibrium and presumably also
primordial black holes by further gravitational collapse.

We use the natural unit of $\hbar = c = 1$ throughout the present work
 unless otherwise stated.
Moreover, we express cosmic temperature in eV unit taking
the Boltzmann constant $k_B = 1$.

After web-submission of the original version to this work,
we were informed, and became aware, of past works
\cite{caldi-chodos}, \cite{kapusta}, \cite{chodos-cooper}
related to the present work.
The main difference of the present work from others is 
that we consider all weak interactions
of charge neutral fermions, which leads to the attractive force
between left-handed neutrino and neutron.
Other works concentrates on neutrino interaction alone, and
for condensate formation are forced to
extend the standard particle physics model.
We insist on the standard model with finite neutrino masses.

\section
{\bf Neutrino-neutron pair condensate}

\subsection{Neutrino interaction}

\begin{figure*}[htbp]
 \begin{center}
 \centerline{\includegraphics{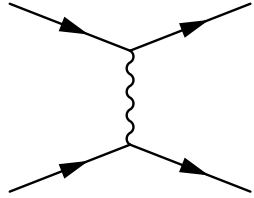}} \hspace*{\fill}
   \caption{
Scattering of $\nu_L$ off  left-handed neutron  via t-channel Z-exchange.
The wavy one is for $Z$ boson.
}
   \label {nu scattering}
 \end{center} 
\end{figure*}

Consider the ideal gas consisting of three left-handed neutrinos in mass eigenstates
denoted by $\nu^i_L\,, i = 1,2,3$ (with $L, R= (1\mp \gamma_5)/2$ the chiral projection)
interacting with other fermions, 
in the early universe at temperatures above 1 MeV  \cite{cosmology}
and below the electroweak phase transition around 250 GeV.
We look for attractive weak forces between a neutral fermion and
other fermions.
Attractive charged particle interaction is not appropriate for our purpose due to
that the phase coherence necessary towards condensate formation
is likely to be destroyed by ambient photons interacting with charged particles.
Left-handed neutrino is the major charge-neutral
 constituent in this universe, and we start to study their interaction.
Neutrinos frequently interact in the early universe at  temperatures above
1 MeV, and their aggregate is well described by
an ideal gas of temperature $T$ with zero chemical potential.
The self-interaction among $\nu_L$'s occurs by Z-boson exchange.
The Higgs exchange interaction mixes $\nu_L$ to the right-handed $\nu_R$,
but their interaction occurs much less frequently.
The Higgs-exchange interaction is thus irrelevant to the present discussion.
The potential of Z-exchange interaction in the coordinate space is given by
\begin{eqnarray}
&&
- \frac{g_Z^2}{32 \pi} \frac{e^{-m_Z r}}{r} 
\left( u_3^{\dagger} u_1 u_4^{\dagger} u_2 - 
u_3^{\dagger}\vec{\sigma} u_1\cdot u_4^{\dagger} \vec{\sigma}  u_2
\right)
\,,
\label {z-exchange amp}
\end{eqnarray}
in terms of two-component spinor wave functions of neutrinos, 
$u_i(p_i)\,, i=1 \sim 4$ with $p_i$ 4-momenta of neutrinos.
This amplitude arises from t-channel Z-boson exchange
in the scattering $\nu(p_1) + \nu(p_2) \rightarrow \nu(p_3) + \nu(p_4)$,
its diagram similar to Fig(\ref{nu scattering}) with neutron replaced by $\nu_L$.

To see the spin-dependence of Z-boson exchange interaction, 
it is convenient to exchange wave functions,
$u_1$ and $u_4^{\dagger}$, using the identity,
\begin{eqnarray}
&&
(u_1 u_4^{\dagger} )_{ij}= \frac{1}{2} \left( u_4^{\dagger} u_1 
+  u_4^{\dagger}\vec{\sigma} u_1 \cdot \vec{\sigma}
\right)_{ij}
\,.
\end{eqnarray}
The sign change due to Fermi-Dirac statistics should be superimposed further.
Two neutrino states in the initial and the final states are
$u_1 u_2$ and $u_3 u_4$, which may be decomposed into
projected states of total spin eigenvalues, singlet and triplet, using the formulas,
\begin{eqnarray}
&&
u_{1i}u_{2j} = \sqrt{2} \left( (S_{12} + \vec{V}_{12}\cdot \vec{\sigma} ) ( - i \sigma_2)
\right)_{ij}
\,, \hspace{0.5cm}
S_{12} = \frac{1}{2\sqrt{2}} u_1^T i \sigma_2 u_2
\,, \hspace{0.5cm}
\vec{V}_{12} = \frac{1}{2\sqrt{2}} u_1^T i \sigma_2 \vec{\sigma} u_2
\,.
\end{eqnarray}
The bracket part  $\left(\cdots \right) $ of eq.(\ref{z-exchange amp}) is then equal to
\begin{eqnarray}
&&
- 2 \left( S_{34}^{*} S_{12} + \vec{V}_{34}^{*} \cdot \vec{V}_{12} \right)
\,.
\end{eqnarray}
A positive pre-factor in (\ref{z-exchange amp})
 means repulsion, while a negative one means attraction.
Thus, one  concludes that Z-boson exchange gives repulsive self-interaction 
\cite{caldi-chodos}
both for spin-singlet and triplet, with a strength $+ g_Z^2 e^{-m_Z r}/(16 \pi r )$.

Neutrinos however interact with other fermions;
electron, u-quark and d-quark at QCD epochs around 100 MeV temperature.
The strength of Z-boson exchange interaction among fermion species $f$ is given by
\begin{eqnarray}
&&
\sqrt{g^2+ (g')^2}  Z \cdot \bar{\psi}_f
\gamma \, \left( T_3 L - \sin^2 \theta_w Q_f (L + R) \psi_f
\right)  
\,,
\end{eqnarray}
with $T_3 = \sigma_3/2$ SU(2) quantum numbers and $(L, R)= (1 \mp \gamma_5)/2$.
$\psi_f$ is the field operator of species $f$.
We shall first work out these interaction at epochs after anti-nucleon annihilation
around $\sim $ 20 MeV and prior to nucleo-synthesis at $\sim$ 0.1 MeV, to give
relative interaction weights among
left-handed fermions in surrounding medium
\begin{eqnarray}
&&
p_L\; :\;  \frac{1}{4} - \frac{1}{2} \sin^2 \theta_w 
\,, \hspace{0.5cm}
n_L \; :\;  -  \frac{1}{4}
\,, \hspace{0.5cm}
e_L \; :\; 
- \frac{1}{4}|U_{ie}|^2  + \frac{1}{2} \sin^2 \theta_w
\,, \hspace{0.5cm}
(\nu_i)_L \; :\; \frac{1}{4} 
\,.
\end{eqnarray}
When the product of these is positive, their force is repulsive,
while it is attractive for a negative product.
$\theta_w$ is the weak mixing angle measured to be
$\sin^2 \theta_w \sim 0.238$, and 
$U_{ie}$ is a neutrino mass mixing matrix element of order
$ |U_{ie}|^2 \sim 0.7 $ \cite{pdg}.
Interaction between $\nu_e$ and $ e_L$ 
is mediated by W-exchange as well.
Numerically, these numbers are $0.13\,, - 0.25\,, -0.056 \,, 0.25$, respectively.
Thus, attractive Z-exchange interaction of  left-handed neutron exists between
left-handed neutrino and proton.
It is easy to understand the attractive nature of force between $\nu_L$ and $n_L$,
since they have opposite Z-charges proportional to $T_3= \pm 1/2$,
much like the product sign between unlike charges.
Interaction of left-handed neutron with right-handed electron is also attractive. 
Right-handed neutron does not interact with other chiral fermions to this order
of tree level diagram.

We are thus led to investigate the possibility of
left-handed neutrino-neutron condensate formation.
From the point of uniform condensed state it is 
advantageous to take a total zero momentum pair 
of $(\vec{p}\,, - \vec{p})$ and spin-singlet state.
This poses a problem, however:
since neutrons below 20 MeV are non-relativistic, the fraction of condensation
partner, relativistic $\nu_L$'s,  is  limited to a small portion of phase-space.
This leads us to explore even earlier epoch of abundant quarks and anti-quarks
much above 20 MeV.

Relative Z-couplings of u- and d-quarks are given by
\begin{eqnarray}
&&
u_L \; ; \frac{1}{4} - \frac{1}{3} \sin^2 \theta_w \sim 0.17
\,, \hspace{0.3cm}
d_L \; ; - \frac{1}{4} + \frac{1}{6} \sin^2 \theta_w \sim - 0.21
\,,
\\ &&
u_R \; ;  - \frac{1}{3} \sin^2 \theta_w \sim - 0.079
\,, \hspace{0.3cm}
d_R \; ;  \frac{1}{6} \sin^2 \theta_w \sim 0.040
\,.
\end{eqnarray}
Thus, left-handed neutrinos have attractive interaction with $d_L$ and
$u_R$ (to a smaller strength).
These quarks may be treated as massless at high temperatures.

The effective strength of spin-singlet 
$\nu_L d_L$ interaction at finite temperatures is estimated by
taking a thermal average over neutrinos, to lead to
\begin{eqnarray}
&&
V_{\nu} = -
\frac{g_Z^2}{6\zeta(3)} \frac{1 }{ r} \int_0^{\infty} dx \,
\frac{x \sin (x m_Z r) }{ (x^2+ 1) (e^{x m_Z/T} + 1) }
= \frac{g_Z^2}{6\zeta(3)} \frac{1 }{ r} \Im \int_0^{\infty} dx \, 
\frac{x e^{i x m_Z r}}{x^2+1} \left( \sum_{n= 1}^{\infty} (-1)^{n+1} e^{- n x m_z/T}
\right)
\,.
\end{eqnarray}
The integral may be estimated by the residue at a pole
$x = i$ in the complex plane of $x$,
to give
\begin{eqnarray}
&&
V_{\nu} = - \frac{\pi}{12 \zeta(3)} \frac{g_Z^2}{r} e^{-m_Z r}
\,.
\end{eqnarray}
This is larger than the free-space value by $4\pi^2/3 \zeta(3) \sim 7.3$.

Consider  i-th massive neutrinos with its number density $n_i$.
The energy density of neutrino-neutron pairs at finite densities is given by
\begin{eqnarray}
&&
\sqrt{k^2 +m_i^2} \,n_i + m_n  + V_{\nu} n_i^2
\,.
\label {n dependent potential 2}
\end{eqnarray}
The system is unstable for 
\begin{eqnarray}
&&
n_i >  \frac{ \sqrt{k^2 +m_i^2}}{|V_{\nu}| }
\,, \hspace{0.5cm}
\frac{m_i}{|V_{\nu}| } = 0.22 \times 10^2 \,{\rm MeV}^3 \frac{m_i}{ 50 {\rm meV}}
= 0.16 \times 10^5 {\rm cm}^{-3} \frac{m_i}{ 50 {\rm meV}}
\,.
\end{eqnarray}
Note a much larger typical neutrino number density in the early universe,
\begin{eqnarray}
&&
n_{\nu} = \frac{3 \zeta(3)}{2\pi^2} T^3 \sim 2.4 \times 10^{ 31}\,
{\rm cm}^{-3} (\frac{T}{{\rm MeV}})^3
\,.
\end{eqnarray}

\begin{figure*}[htbp]
 \begin{center}
 \centerline{\includegraphics{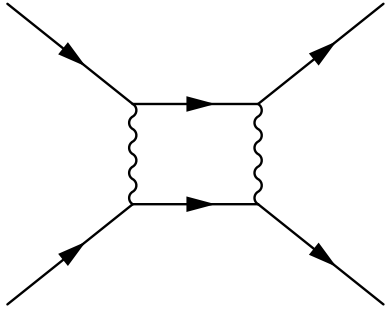}} \hspace*{\fill}
   \caption{
Scattering of $\nu_L$ off neutron via one-loop $W^{\pm}\,, Z$ exchange box diagram.
Straight lines are for $\nu_L$, while
wavy ones are for $W^{\pm}\,, Z$ bosons.
}
   \label {wz box}
 \end{center} 
\end{figure*}

As in the theory of superconductivity \cite{superconductivity}  in which phonon mediated
interaction for a Cooper pair of electrons of momenta, 
$(\vec{p}, -\vec{p})$, gives rise to pair condensation,
one may expect a similar phenomenon of $\nu_L d_L$ super-fluid.
First, let us discuss a bound state of the pair.
Higher order interaction between spin-singlet neutrino pair is given by 2Z and 2W
exchange and there are still higher orders like Fig(\ref{wz box}).
Summing up all these terms gives
\begin{eqnarray}
&&
- \frac{ |V_{\nu}| n_i^2}{1- t_Z - t_W }
\\ &&
t_Z = \frac{3g_Z^2 }{4\pi^2 } \ln \frac{m_Z^2}{m_i^2}
\,, \hspace{0.5cm}
t_W = \frac{3g_W^2 }{4\pi^2 } \ln \frac{m_W^2}{m_i^2}
\end{eqnarray}
We have neglected sub-leading terms proportional to
$m_l^2/m_W^2$ arising from charged leptons of mass $m_l$.
Bound pair state thus appears at 
\begin{eqnarray}
&&
 m_i - \frac{ |V_{\nu} |n_i}{1- t_Z - t_W } < 0
\,,
\end{eqnarray}
 which gives a smallest neutrino mass value for given number density $n_i$.

Pairs made of $\nu_L d_L$ can attract $u_L\,, u_R$ by strong Coulomb force.
It is likely that this develops into bound-pair formation of two neutrals, $\nu_L$ and
left-handed neutron $n_L$.
This hadron formation is a complicated process, since the majority
of internal energy of nucleons is made of QCD gluons.
We shall simply assume that there is no difficulty of
$\nu_L n_L$ bound-pair formation of spin-singlet $(\vec{p}, - \vec{p})$ configuration.

To show that bound $\nu_L n_L $ state pairs actually condensate, 
one further needs to discuss
global and macroscopic features of many pair system, and for this purpose
we analyze the problem using
the Ginzburg-Landau mean field theory and Bogoliubov transformation
\cite{landau} along the same lines of arguments as in superconductivity.

\subsection
{\bf Ginzburg-Landau mean field theory and Bogoliubov transformation}

\begin{figure*}[htbp]
 \begin{center}
 \centerline{\includegraphics{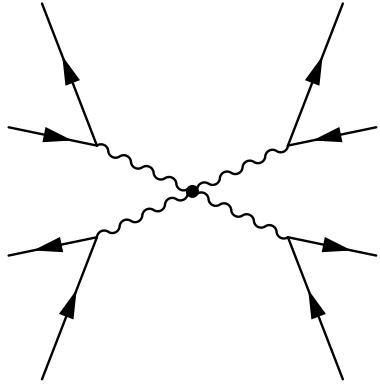}} \hspace*{\fill}
   \caption{
Six $\nu_L$ and two $n_L$ attached to four Z-vertex. 
}
   \label {eight nu}
 \end{center} 
\end{figure*}

We assume a homogeneous density of condensate $n$ and calculate
terms proportional to $n^4$ (term $\propto n^3$ does not occur).
The Feynman diagram for this process is four Z vertex with neutrino-neutron pair 
$\nu_L n_L$ attached to each propagating Z, as shown in Fig(\ref{eight nu}).
One loop diagram of circulating neutrino and other fundamental particles
 coupled to four Z gives
divergent contribution which gives renormalization counter term to
the bare four Z vertex.
In terms of renormalized four Z coupling this contribution gives
repulsive term of strength,
\begin{eqnarray}
&&
\frac{4 G_F}{\sqrt{2} } \frac{g_Z^6}{m_Z^6} n^4 
> 0 \; {\rm for\; singlet}
\,.
\end{eqnarray}
Potential minimum when $m_{\nu}=0$ is at a critical density $n_c$, which
is given by 
$n_c \approx 2\pi m_Z^3/ (\sqrt{96 \zeta(3)} g_Z^3) \sim 0.48 (m_Z/g_Z)^3$, 
close to electroweak phase transition point.
More precisely,
\begin{eqnarray}
&&
n_c = \frac{2\pi }{\sqrt{96 \zeta(3)}} (\frac{m_Z}{g_Z})^3
\left( 1 + \frac{\sqrt{16} m_{\nu} }{G_F}  (\frac{g_Z}{m_Z})^3
\right)
\,.
\end{eqnarray}
Condensate energy density is given by
\begin{eqnarray}
&&
2 m_{\nu} n_c - \frac{3  G_F}{ 4 \sqrt{2}} n_c^2
\left( 1 + \frac{\sqrt{32} m_{\nu} }{G_F}  (\frac{g_Z}{m_Z})^3
\right)
\label {condensate energy}
\end{eqnarray}

For later application we recapitulate the operator form of
singlet pair hamiltonian density,
\begin{eqnarray}
&&
H_{\nu} = 2 m_{\nu} n + \frac{3 G_F}{4\sqrt{2} } 
\left( - n^2 + 16 ( \frac{g_Z }{ m_Z})^6 n^4
\right)
\,.
\end{eqnarray}

Due to instability of bound state formation the ground state is radically changed
from the usual one at zero temperature, and one needs to find a new ground state, 
which is done by using a Bogoliubov transformation.
The Bogoliubov transformation violates lepton-number, and is introduced by
a linear combination of annihilation and creation operators of definite momentum
$\vec{k}$,
\begin{eqnarray}
&&
\beta_k = u_k b_k - v_k b_{-k}^{\dagger}
\,, \hspace{0.5cm}
\beta_k^{\dagger} = u_k b_k^{\dagger} - v_k b_{-k}
\,.
\end{eqnarray}
with real functions, $u_k\,, v_k$, to be determined.
We anticipate that the state $| \tilde{0} \rangle $ 
defined by $\beta_k | \tilde{0} \rangle =0$ for all $\vec{k}$ 
is the ground state made of condensates, and
$\beta_k^{\dagger} | \tilde{0} \rangle$
is a quasi-particle state of single neutrino in super-fluid.

Anti-commutation
relations, $\{ \beta_k\,,  \beta_{k'}^{\dagger} \} = \delta_{k,k'} \,, 
\{\beta_k\,, \beta_k' \} = 0$,
requires
\begin{eqnarray}
&&
u_k^2 + v_k^2 =1
\,, \hspace{0.5cm}
u_k v_{-k'} + v_k u_{-k'} = 0 
\,.
\end{eqnarray}
The latter condition is readily satisfied by $u_{-k} = u_k\,, v_{-k} = - v_k$.
With this property, the inversion is given by
\begin{eqnarray}
&&
b_k = u_k \beta_k + v_k \beta_{-k}^{\dagger}
\,, \hspace{0.5cm}
b_k^{\dagger} = u_k \beta_k^{\dagger} + v_k \beta_{-k}
\,.
\end{eqnarray}

In the presence of condensate 
one defines the order parameter
$\Psi = \sqrt{n_c} + \sum_k (b_k + b_k^{\dagger})$ and
the number density operator $\tilde{n} = \Psi^{\dagger} \Psi$.
Introducing a hermitian operator $A$, we derive relations,
\begin{eqnarray}
&&
A_k =  (u_k - v_k) (\beta_k + \beta_k^{\dagger} )
\,,
\\ &&
\tilde{n} = n_c + \sum_k  A_k^2 + \sqrt{n}_c \sum_k  A_k
\,,
\\ &&
\langle \tilde{0} | n | \tilde{0} \rangle = n_c + \sum_k (u_k - v_k)^2 f_k
= n_c +  \sum_k (u_k^2 + v_k^2\,) f_k = 2 n_c
\,,
\\ &&
\sum_k f_k = \frac{d^3 k}{(2\pi)^3} f_k = n_c
\,,
\end{eqnarray}
with $f_k$ the momentum distribution function assumed spatially homogeneous.
The distribution function here refers to neutrinos within a condensate,
not to be confused with the thermal distribution in surrounding medium.
This is a decomposition of number operator into the mean field value of
condensates and elementary excitation (or quasi-particle).

We give two important examples of momentum distribution:
the first is
Fermi-Dirac distribution at finite temperatures applicable in the early stage
of condensate formation,
\begin{eqnarray}
&&
f_k = \frac{1}{e^{\sqrt{k^2 + m_{\nu}^2}/T} + 1} 
\approx \frac{1}{e^{k/T} + 1} 
\,, \hspace{0.5cm}
n_c = \int \frac{ d^3 k }{(2\pi)^3} f_k
\approx \frac{\zeta(3)}{4\pi^2}T^3  \sim 0.046\, T^3
\end{eqnarray}
The other is degenerate distribution in beta equilibrium,
$n + (\nu_e)_L \leftrightarrow p + e_L$, applicable in the late stage
of condensate evolution:
\begin{eqnarray}
&&
f_k = \theta( k_F -  |\vec{k}| )
\,, \hspace{0.5cm}
n_c = \int \frac{ d^3 k }{(2\pi)^3} f_k = \frac{ 1}{ 6 \pi^2}k_F^3 
\,.
\end{eqnarray}

We state two theorems and results on super-fluid state of
neutrino singlet pair.

{\bf Theorem} \lromn1
\hspace{0.5cm} 
{\bf Emergence of super-fluid}

First, note the following identity:
\begin{eqnarray}
&&
[ A^2 \,, \beta_k ] = 0
\end{eqnarray}
We use normal-ordered operator product as in the Wick theorem
in quantum field theory, which gives contracted factor,
\begin{eqnarray}
&&
\langle \tilde{0} | A^2 | \tilde{0} \rangle = \sum_k (u_k^2 + v_k^2) = n_c
\end{eqnarray}
The following relations then hold:
\begin{eqnarray}
&&
\langle \tilde{0} | (A^2)^p | \tilde{0} \rangle = n_c^p
\\ &&
\langle \tilde{0} |\beta_1 \cdots \beta_s (A^2)^p 
\beta_s^{\dagger} \cdots \beta_1^{\dagger} | \tilde{0} \rangle = n_c^p
\end{eqnarray}
for $1\,, 2\,, \cdots l$ taken all different.
Expectation values of any function of number operator $f(n)$
are then calculated as
\begin{eqnarray}
&&
\langle \tilde{0} |\beta_l \cdots \beta_s f(n) 
\beta_s^{\dagger} \cdots \beta_1^{\dagger} | \tilde{0} \rangle = 
\frac{1}{2} \left( f(n_c^+) + f(n_c^-)
\right)
\,, \hspace{0.5cm}
n_c^{\pm} = 2n_c \pm \sqrt{n_c} A
\,.
\end{eqnarray}
This implies that macroscopic event involving neutrinos in condensate
has no resistance, hence it proves the super-fluid nature of condensates.

These expectation values are independent of number of neutrinos involved,
and this suggests that neutrinos in condensate behave as  massless.
We shall prove this in the following theorem.

{\bf Theorem} \lromn2
\hspace{0.5cm} 
{\bf Condensate energy and massless quasi-particle}

We consider effective hamiltonian (actually density)  to order $n^4$, 
to derive operator form,
\begin{eqnarray}
&&
H_{\nu} = m_{\nu} n + \frac{3 G_F}{ 4 \sqrt{2}} (- n^2 + 16 (\frac{g_Z}{m_Z} )^6 n^4 )
\end{eqnarray}
Following the method of Theorem \lromn1, we calculate energies of
condensate  and many quasi-particle state:
\begin{eqnarray}
&&
E(\tilde{0}) =
\langle \tilde{0} | H_{\nu} | \tilde{0} \rangle = 
2 m_{\nu} n_c + \frac{3 G_F}{ 4 \sqrt{2}} (- c_2 n_c^2 + 16 c_4 (\frac{g_Z}{m_Z} )^6 n_c^4 )
\\ &&
E(\tilde{s}) =
\langle \tilde{0} |\beta_s\cdots \beta_1
 H_{\nu} \beta_1^{\dagger} \cdots \beta_s^{\dagger} | \tilde{0} \rangle = 
\langle \tilde{0} | H_{\nu} | \tilde{0} \rangle = E(\tilde{0})
\; {\rm for \;} (1,2, \cdots s) \; {\rm all\; different} 
\end{eqnarray}
with $c_2 =5\,,c_4 = 41 $.

This energy density has two extrema for $m_{\nu} > 0$,
a local maximum near $n_c=0$ and a global minimum $n= n_*$, with
\begin{eqnarray}
&&
n_* = \sqrt{\frac{c_2}{32 c_4}} (\frac{m_Z}{g_Z}  )^3 
\,, \hspace{0.5cm}
E(\tilde{0})_{n_c = n_*} \approx  - \frac{3 c_2^2}{16^2 \sqrt{2} c_4} G_F  
( \frac{m_Z}{g_Z} )^6
\,. \label {critical n}
\end{eqnarray}
This state is highly degenerate due to excitation of  massless neutrinos
of arbitrary numbers.

\section
{\bf Cosmological evolution of pair condensates}

The critical temperature $T_c$ is estimated by equating 
$n_*$ in eq.(\ref {critical n}) to thermal number density of neutrino
$3 \zeta(3)T^3/(4 \pi^2  ) $ (for one spin component $\nu_L$
not counting $\bar{\nu_L}$), to derive
\begin{eqnarray}
&&
T_c = ( \frac{  \pi^2 c_2}{16 \zeta(3) c_4})^{1/3} \frac{m_Z}{g_Z} 
\sim 0.40 \, \frac{m_Z}{g_Z}  \sim 71 \,{\rm GeV}
\,.
\end{eqnarray}
Thus, the critical temperature is less than temperature of electroweak phase transition.
Below $T_c$ the instability develops towards pair condensation.

\subsection
{Super-fluid phases and condensate number density}

Important properties of super-fluids made of $\nu_L n_L$ pairs are
described by the space-time dependent order parameter 
$\Phi( \vec{r}, t) = \nu_L( \vec{r}, t) n_L( \vec{r}, t)$ in spin-singlet combination,
just as by Cooper pair in the ordinary superconductor.
Main differences from there are charges and masses of constituents.
We shall ignore  space dependence of order parameter, and focus on
cosmological time evolution.
Order parameter of condensate is decomposed as a product of
absolute magnitude $\sqrt{n_S} > 0$ and the phase function.

Following  \cite{superconductivity}, we study the super-current
across a boundary (called junction
in solid state physics) between two super-fluid phases of pair condensates.
We consider a plane boundary and two super-fluid phases
given by  $\Psi_{1}\,, \Psi_2$,
left and right super-fluids with different chemical potentials.
The relevant coupled equations are
\begin{eqnarray}
&&
i \left( \frac{\partial \Psi_1}{\partial t} + 3 \frac{\dot{a}}{a}  \Psi_1 \right) = 
\mu_1 \Psi_1 + {\cal T} \Psi_2
\,,
\\ &&
i \left( \frac{\partial \Psi_2}{\partial t} + 3 \frac{\dot{a}}{a}  \Psi_2 \right) = 
\mu_2 \Psi_2 + {\cal T} \Psi_1
\,,
\end{eqnarray}
with $\mu_{1,2}$ chemical potentials.
For simplicity time-reversal invariance is assumed, which makes  
the transition amplitude ${\cal T}$ real \cite{landau}.
These coupled equations are identical to the junction equations in ordinary
super-conductivity except the effect of cosmic expansion $\propto 3\dot{a}/a$ terms
and time dependences of $\mu_i,\,, {\cal T}$.
One derives from these coupled equations,
\begin{eqnarray}
&&
(\frac{\partial }{\partial t} +  3 \frac{\dot{a}}{a})|\Psi_2|^2 = 
 - 2{\cal T} \sin (\chi_2 - \chi_1)  |\Psi_2|^2 
\,,
\label {condensate evolution eq}
\\ &&
(\frac{\partial }{\partial t} +  3 \frac{\dot{a}}{a}) (|\Psi_1|^2 + |\Psi_2|^2 ) = 0
\,,
\end{eqnarray}
with $|\Psi_1|^2  = |\Psi_2|^2 = n_S$, recovering the momentum fraction factor.
On the other hand, the equation for $ (\partial_t + 3 \dot{a}/a ) (\Psi_{1} \Psi_{2})$
leads to
\begin{eqnarray}
&&
\partial_t (\chi_1 - \chi_2) = - (\mu_1 - \mu_2) \equiv \delta \mu_{12}
\,,
\\ &&
(\frac{\partial }{\partial t} +  3 \frac{\dot{a}}{a})  n_S = 
 - 2{\cal T} \left( \sin \int^t dt' \delta \mu_{12}(t') \right)  n_S 
\,.
\end{eqnarray}

Difference from  equation of stable decoupled matter is obvious:
there exists continuous flow in and out from adjacent super-fluids
if both ${\cal T}\,, \delta \mu$ are non-vanishing.
This gives rise to the logarithm of number density decreasing as
the third power times correction given by an integral,
hence the number density of super-fluid remains constant as
cosmic time increases.
The equation for super-fluid number density is formally solved, to give
\begin{eqnarray}
&&
n_S(t) = n_S(t_c) \exp[  -2  \int_{t_c}^{t} dt_1
{\cal T}(t_1) (\frac{a(t_1)}{a(t)})^3  
\left( \sin \int^{t_1}_{t_c} dt'' \delta \mu(t'') \right) ]
\,.
\label {super-fluid density: time evolution}
\end{eqnarray}
It may be better to give this solution in terms of cosmic temperature;
\begin{eqnarray}
&&
(\frac{a(t_1)}{a(t)})^3 = (\frac{T}{T_1})^3  
\,,
\end{eqnarray}
and $dt \propto dT/T^3 \,, \propto dT/T^{5/2}$ depending on
radiation- or matter-dominated epoch.

Chemical potentials in super-fluids are identified as the
free neutrino energy written as $\mu = \sqrt{p^2+ m_{\nu}^2}$
plus interaction term of order $G_F n$, both real numbers.
In two adjacent phases interaction are identical, but
neutrino masses may be different.
The result of time evolving super-fluid number density derived above 
depends on difference of chemical potential in adjacent phases.
Thus,  there are three different types of boundaries for three massive neutrinos.
We are interested in relativistic neutrinos in the early epoch, hence
$\sqrt{p^2+ m_{\nu}^2} \sim p + m_{\nu}^2/2p $,
giving $\delta \mu_{ij} = (m_i^2- m_j^2)/2p$.
We shall be content with an approximation in which neutrino momentum
here is replaced by its thermal average:
\begin{eqnarray}
&&
\frac{1}{\bar{p}} = \frac{4 \pi^2 \zeta(3) }{3 T } \int_0^{\infty} dx\, \frac{x}{e^x+ 1} 
= \frac{ \pi^4 \zeta(3) }{9 T } \sim \frac{19.5 }{T}
\,.
\end{eqnarray}
This gives
\begin{eqnarray}
&&
\int^{t_1}_{t_c} dt'' \delta \mu(t'') = \frac{ \pi^4 \zeta(3)  }{18 c } \int_T^{T_c}
 dT' \frac{ \delta m_{ij}^2 M_{{\rm pl}} }{ (T')^4} = \frac{ \pi^4 \zeta(3)  }{54 c } 
\delta m_{ij}^2 M_{{\rm pl}} (\frac{1}{T^3} - \frac{1}{T_c^3} )
\,,
\\ &&
\frac{\delta m_{ij}^2 M_{{\rm pl}} }{T^3}
	\sim 1.2 \times 10^{-6}
 \frac{\delta m_{ij}^2 }{ (10 {\rm meV})^2} (\frac{10 {\rm GeV}}{T})^3
\,, 
\end{eqnarray}
with coefficient $ \frac{ \pi^4 \zeta(3)  }{54 c }$ 
of order 0.1 at 1 Gev region of temperature.

The transmission matrix element ${\cal T}$ 
is given by real part of forward scattering amplitude times number density
of scattered target in the boundary wall.
For a simple estimate we shall replace the real part by the imaginary 
part or absorptive part, which is expressed in terms of total cross section
by the optical theorem.
With this assumption we estimate 
\begin{eqnarray}
&&
{\cal T}(T) \approx \sigma_{{\rm tot}}  n =
\frac{ G_F^2 T^5} {2\pi^3} \int_{0}^{\infty} dx\, \frac{x^4}{e^{x}+ 1}
= \frac{ 45 \zeta(5) G_F^2 T^5} {4\pi^3} \sim 8.5 \, G_F^2 T^5
\,,
\\ &&
dt_1
{\cal T}(t_1) (\frac{a(t_1)}{a(t)})^3   = - G_F^2 
dT_1 \, \frac{M_{{\rm pl}} T^3}{  T_1}
 \frac{45 \zeta(5)}{4 \pi^3 c}
\approx - 5.75 \times 10^{ 11} \,
(\frac{T}{10 {\rm GeV}})^3 \frac{dT_1}{T_1} 
\,.
\end{eqnarray}

The exponent of $n_S(t)/n_S(t_c)$ exhibits an interesting behavior:
above a few times 100 MeV temperature region  it is a large number
independent of temperature.
It starts to rapidly oscillate when the argument of sinusoidal function
 becomes of order unity, which
is around 50 $\sim $ 100  MeV. The amplitude is of order $10^5 (T/ 50 {\rm MeV} )^3$
or larger.
The oscillation ends around 10 MeV prior to the onset time of nucleo-synthesis,
decreasing rapidly,
and a large constant number density $n_S(t_c)$ remains at later epochs.

We took here a junction between two super-fluids, but
a junction between super- and normal-phases can be dealt with
in a similar way.

\begin{figure*}[htbp]
 \begin{center}
 \epsfxsize=0.6\textwidth
 \centerline{\epsfbox{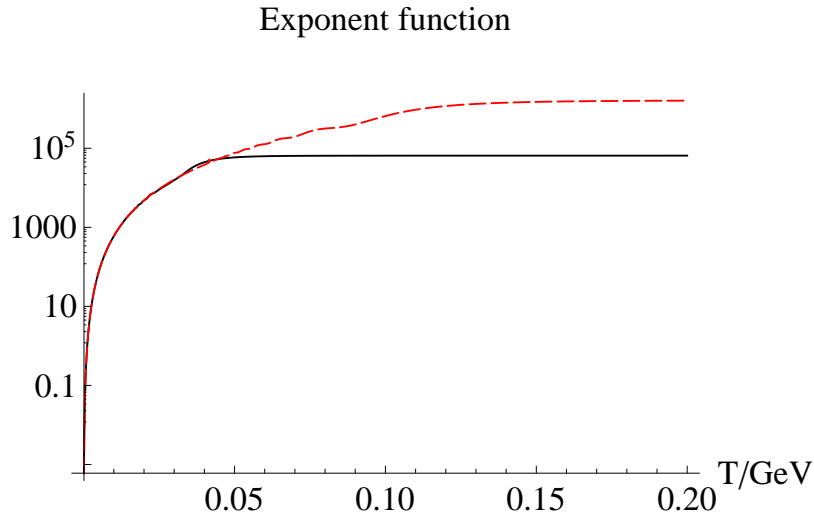}} \hspace*{\fill}\vspace*{1cm}
   \caption{
Exponent function $f(T\,, T_c)$  of super-fluid number density
given by (\ref{exponent function}).
Parameter $\delta^2 m_{ij} $ chosen is
$10\, {\rm meV}^2$ in solid black and $250 \, {\rm meV}^2$
in dashed red.
}
   \label {exponent function fig}
 \end{center} 
\end{figure*}

\begin{figure*}[htbp]
 \begin{center}
 \epsfxsize=0.6\textwidth
 \centerline{\epsfbox{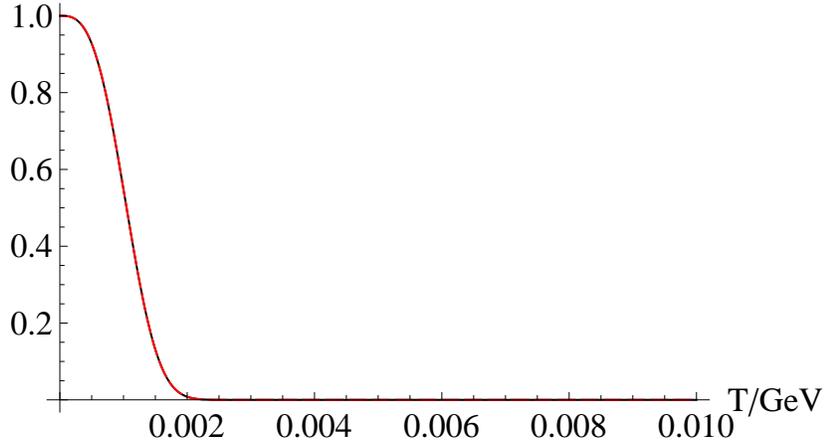}} \hspace*{\fill}\vspace*{1cm}
   \caption{
Super-fluid number density $n_S(T) /n_S(T_c)$
corresponding to the same parameter choice of Fig(\ref{exponent function fig}).
Two parameter $\delta^2 m_{ij} $ cases of
$10\, {\rm meV}^2$  and $250 \, {\rm meV}^2$
give nearly identical $n_s(T)$ in thie temperature range with this calculation
resolution.
Note a smaller temperature range than in Fig(\ref{exponent function fig}).
}
   \label {super-fluid number density}
 \end{center} 
\end{figure*}

We illustrate the behavior of exponent function in Fig(\ref{exponent function fig}).
The formula we used for this is the exponent function,
\begin{eqnarray}
&&
- \frac{45 \zeta(5) G_F^2 M_{{\rm pl}}  }{2 \pi^3 c}  T^3
\int_{T}^{T_c} \, \frac{ dT_1 }{  T_1}
\sin \left( \frac{ \pi^4 \zeta(3)  }{54 c } \delta m_{ij}^2 M_{{\rm pl}}
(\frac{1}{T_1^3}- \frac{1}{T_c^3} )
\right)
\\ &&
\hspace*{-0.5cm}
\sim - 1.15 \times 10^{9} (\frac{T}{1 {\rm GeV}})^3 
\int_{T}^{T_c} \, \frac{ dT_1 }{  T_1} \sin
\left( 1.7 \times 10^{ -4} \frac{\delta m_{ij}^2 }{ (10 {\rm meV})^2}  
(\frac{ 1 {\rm GeV}}{T_1})^3 - (T_1 \rightarrow T_c)
\right)
\equiv - f(T\,, T_c)
\,.
\label {exponent function}
\end{eqnarray}
The corresponding super-fluid number density $n_S(T) /n_S(T_c)$
is shown in Fig(\ref{super-fluid number density}).

What would be the mass distribution of neutrino-neutron condensates ?
Super-fluid formation occurs in the  temperature range
from $T_c$  till around 50 MeV.
Horizon sizes $R$ during this interval are possible sizes of super-fluid phases,
thus ranging in $R = 4  \sim 4 \times 10^{6}$ cm,
whose mass amounts to a fraction of $m_n \times 10^{54}$, a value
smaller than the solar mass, but larger than the earth mass.
Condensates, however, attract each other and they may further grow
till ordinary neutron star of mass, a few times the solar mass.

Time evolution of pair condensate as described above
 is valid only when neither accretion nor leakage of pair condensate occurs.
But as discussed in the next subsection, accretion is likely to proceed.

Before closing this subsection, it is instructive to discuss why 
electron-proton plasma at similar epochs of early universe does not produce
condensate.
Stronger Coulomb attraction between electron and proton might
suggest condensate of hydrogen atoms.
But this stronger interaction produces abundant photons,
to create a plasma state including thermal photons which
 block condensate formation by destroying a necessary phase coherence.
In the neutrino-neutron case mediating Z-boson is energetically impossible to be
produced after the electroweak phase transition such that
no blocking occurs towards neutrino pair condensation.

\subsection
{Accretion of other fermions}

Note first that the ratio of left-handed neutrino density in super-fluid phase to that
in normal phase increases rapidly, with temperature power $T^{-3}$.
It thus seems that leakage of left-handed neutrinos dominantly occurs from over-dense 
super-state to normal state.
However, accretion of other fermions, $p_L\,, e_L\,, e_R\,, \nu_L$ having
attractive interaction with $n_L$, 
may balance against this neutrino leakage.
We consider for definiteness
 the epoch of cosmic temperature in the range $20$ MeV $\sim$ down 
to the onset epoch of nucleo-synthesis $\sim 1 $MeV.

Consider a neutrino pair condensate and calculate
effective interaction hamiltonian of $p, n, e, \nu$ in surrounding thermal bath.
Extending the previous calculation, one finds the effective hamiltonian is given by
\begin{eqnarray}
&&
H_{a} =  \frac{3 G_F}{4\sqrt{2} } 
\left( - \sum_{ij} n_i n_j + 16 ( \frac{g_Z }{ m_Z})^6 \sum_{ijkl} n_i n_j n_k n_l
\right)
\,,
\label {accretion hamiltonian}
\end{eqnarray}
where $n_i\,, i = p,n,e,\nu$ are thermal number densities, and
 their weight factors relative to neutrino $\xi_i$ are given by
$\xi_p = 1- 2 \sin^2\theta_w\,, \xi_n = -1\,, \xi_e = 4 |U_{ei}|^2 - 2\,, \xi_{\nu}= 1$.
Hence the bracket quantity in eq.(\ref{accretion hamiltonian}) is
\begin{eqnarray}
&&
- \sum_{ij} n_i n_j + 16 ( \frac{g_Z }{ m_Z})^6 \sum_{ijkl} n_i n_j n_k n_l
= - (\sum_i \xi_i)^2 n_{\nu}^2 + 16 ( \frac{g_Z }{ m_Z})^6 (\sum_i \xi_i)^4 n_{\nu}^4
\,,
\end{eqnarray}
with $n_{\nu} \approx 3\zeta(3) T^3/(4\pi^2) \sim 0.137 T^3$
using the number density appropriate to the early stage of condensate formation.
Thus one may estimate the temperature $T_a$ at which accretion starts:
\begin{eqnarray}
&&
T_a = \left( \frac{ 1}{ 16(\sum_i \xi_i)^2 } \right)^{1/6} \frac{m_Z}{g_Z} \sim 
69 \, {\rm GeV}
\,,
\end{eqnarray}
with $\sum_i \xi_i = 5 - 2 \sin^2 \theta_w \sim 4.52$ adding
three neutrino species, and using $ m_Z/g_Z \sim 180 {\rm GeV}$.
Thus, accretion of surrounding fermions becomes possible slightly below, 
but very close to, the critical temperature $T_c$.

Accretion  is, in its detail, likely to proceed in a different way due to asymmetric
accumulation of proton and electron and
stronger Coulomb attraction.
But inside condensates, it is likely that beta equilibrium is realized by
 inverse beta process, $p+e \leftrightarrow n + \nu_e$.
The accretion starts at $\sim 100 $MeV, when super-fluid settles to the
number density $n(T_i)$ with $T_i = 100 \sim 0.1 $GeV,
and proceeds until number densities of $p,n,e$ inside become comparable to their
thermal density outside the bubble.

Higgs boson exchange may also contribute to the attractive force: one can use
the following Fierz identity for four component  Dirac field,
\begin{eqnarray}
&&
\bar{u}_1 (1- \gamma_5) u_2 \, \bar{u}_3 (1- \gamma_5) u _4 =  
- 2 \left(
\bar{u}_1 \gamma^{\mu} \frac{1+ \gamma_5}{2} u_4 \, 
\bar{u}_3  \gamma_{\mu} \frac{1+ \gamma_5}{2}u _2
\right)
\nonumber \\ &&
\sim 2 u_1^{\dagger} \vec{S} u_4 \cdot u_3^{\dagger} \vec{S} u_2 + \cdots
\end{eqnarray}
The scenario based on Higgs-exchange interaction is pursed in \cite{kapusta},
but it works only for Dirac type of neutrinos.
Moreover, the interaction strength is usually much  weaker.
By introducing a hypothetical Higgs-like boson,
the paper \cite{chodos-cooper} explored a model for $\nu_L-$ pair condensate.
We however  insist on the standard electroweak theory.

The created aggregate of beta-equilibrium
may be called primordial neutron star (PNS), but its internal structure,
in particular, the stratified onion skin structure, \cite{shapiro-teukolsky}
 of ordinary neutron stars is not realized, since ordinary neutron stars are
gravitationally bound.
Gravity affects $\nu_L n_L$ condensates and PNS's at later epochs, 
and may form bosonic stars   \cite{boson star} 
of mass $\sim 10^{13}$ gr, made of $\nu_L n_L$ condensates
this time, while   axion stars have a different mass, $\sim
10^{28} {\rm gr} \, (m_a/\mu {\rm eV})^{-1}$,
\cite{takasugi-yoshim}, \cite{axion star 2}.
Important corrections may arise due to that   $\nu_L n_L$ pairs are 
more fragile than axions.

Number and energy density densities of $\nu_L n_L$
condensates stays constant with cosmological
time evolution, hence they behave as dark energy.
On the other hand, the energy density of PNS may be dominated by
mass density of moving nucleons in beta equilibrium, 
hence they are classified and behave as dark matter.
The criterion of either dark energy or dark matter
is in the form of energy-momentum tensor,
its simplest criterion given by the ratio of pressure $p$ and energy density $\rho$
in isotropic medium $w = p/\rho$ (called the equation of state factor \cite{cosmology}):
$w=-1$ for dark energy and $w=0$ for cold dark matter made of non-relativistic
particles.
Masses of individual PNS's may vary, although they have a mass range
of astrophysical scale, quite unlike a definite mass assigned 
to particle physics dark matter; WIMP and axion.
It is an attractive idea that
two different forms of constituents of universe originate from
a single event, since the dark energy and the dark matter are roughly
comparable energetically at the present epoch.
It must be verified, however, that the major fraction of neutrinos at earlier epochs
does not proceed to formation on 
$\nu_L n_L$ condensates to over-close the universe at late epochs.

\section
{Summary and Outlook}

We studied the nature of neutrino interaction
in the standard electroweak theory, and found that the force is attractive
between left-handed neutrino and neutron.
This led us to investigate the possibility of neutrino-neutron pair condensate
using the standard tools in statistical mechanics and condensed matter physics,
the Ginzburg-Landau mean field theory and the Bogoliubov transformation.
Results suggest that super-fluid formation is likely to occur
in much the same way as condensate formation
of the electron Cooper pair in superconductive metals.
It is likely that the early universe at temperatures below 100 GeV
is made of a mixed phase of ideal gas and a small fraction of
super-fluid pair condensates.

Accretion of other fermions, proton and
neutrino, from surrounding cosmic medium
starts immediately after neutrino-neutron pair condensate
formation, and primordial neutron stars (PNS) in beta equilibrium
 of varying masses are  formed helped by Coulomb attraction
between proton and electron.
An interesting scenario to late-time cosmology is that
condensed neutrino-neutron pairs  provide the dark energy
and further evolved PNS becomes the dark matter.
Estimate of PNS dark matter not to over-close the universe is yet to be
established, and PNS mass spectrum is crucial to determine
their search strategy.
These important issues are left to future works.

It is still premature to judge whether the suggested scenario in the present work
is viable or not. Nevertheless,
we may list a few items on how the proposed scenario may be verified in future.
Detection of relic neutrino may directly probe the momentum distribution
of condensed neutrino pairs.
The idea of detecting relic neutrino proposed in \cite{relic nu detection} is to use
atomic de-excitation, radiative neutrino pair emission,
$| i \rangle \rightarrow |f \rangle + \gamma + \nu_i \bar{\nu_j}$
from macro-coherently  state $|i \rangle$ excited by high-quality lasers,
and to experimentally investigate
how ambient relic neutrinos Pauli-block neutrino pair emission
such that this is reflected in observed photon energy spectrum,
or its accumulated parity violating magnetization that 
helps to reject possible QED backgrounds \cite{mag renp}.
A large neutrino degeneracy created by the initial pair condensate formation
gives a large deviation from
the thermal distribution of zero chemical potential.

Discovery of gravitational wave from black hole 
and neutron star merger enhanced the possibility that some day
one may be able to measure the mass spectrum of relevant neutron stars.
In this respect,
simultaneous detection of gravitational wave and neutrino from
mergers of black hole and PNS   gives important insights
on our scenario.
Crucial question on observable consequences of super-fluid neutrinos
in PNS remains to be studied, but
the proposal may be verified by high statistics data of neutrino burst at PNS formation.

\vspace{1cm}
 {\bf Acknowledgements}

The author would like to thank A. Chodos who kindly
pointed out a sign mistake in the original version of this work.
This research was partially
 supported by Grant-in-Aid   21K03575   from the Japanese
 Ministry of Education, Culture, Sports, Science, and Technology.


\begin{thebibliography}{99}

\bibitem{hayashi}
C. Hayashi,
Progr. Theor.Phys. {\bf 5}, 224 (1950).


\bibitem{superconductivity} 
J.M. Ziman, 
{\it Principles of the Theory of Solids},
Chapter 11, Cambridge University Press (1979).

\bibitem{caldi-chodos}
D.G. Caldi and A. Chodos,
hep-ph/9903416;
{\it Cosmological Neutrino Condensates}.
It was  pointed out by A. Chodos that the original version of the present work
contains a sign mistake on the Z-exchange self-interaction between
two $\nu_L$'s:
the Z-boson exchange interaction gives
repulsion between two left-handed $\nu_L$'s, as explained
in this pioneering work.

\bibitem{kapusta}
J.I. Kapusta,
Phys.Rev.Lett.{\bf 93}, 251801 (2004).

\bibitem{chodos-cooper}
A. Chodos and F. Cooper,
Phys.Rev.{\bf D102}, 113003 (2020).



\bibitem{cosmology}
A standard textbook of modern cosmology is
S. Weinberg, {\it Cosmology},
Oxford University Press, New York (2008).



\bibitem{landau}
E.M. Lifshitz and L.P. Pitaevskii,
{\it Course of Theoretical Physics Volume 9 Statistical Physics: Part 2},
3rd edition. Chapter \lromn5, Pergamon Press (1980).




\bibitem{shapiro-teukolsky}
S.L. Shapiro and S.A. Teukolsky,
{\it Black holes, White dwarfs, and Neutron stars},
John Wiley and Sons (New York), (1983).


\bibitem{boson star}
 R. Ruffini and S. Bonozzola, 
Phys. Rev. {\bf 187}, 1767 (1969).

\bibitem{takasugi-yoshim}
E. Takasugi and M. Yoshimura,
Z.Phys.C {\bf 26},  241 (1984).

\bibitem{axion star 2}
E. W. Kolb and I. I. Tkachev,  Phys. Rev. Lett. {\bf 71}, 3051
(1993).



\bibitem{pdg}
For a review of measurements on the weak mixing angle and
 neutrino mass mixing matrix, see
Particle Data Group Collaboration, M. Tanabashi
et al., Phys. Rev. {\bf D98}, 030001 (2018).



\bibitem{relic nu detection}
M. Yoshimura, N. Sasao, and M. Tanaka,
Phys. Rev.{\bf D90}, 013022 (2014);
{\it Experimental method of detecting relic neutrino by atomic de-excitation},
arXiv; 1403.6546[hep-ph].

\bibitem{mag renp}
H. Hara, A. Yoshimi, and M. Yoshimura,
Phys.Rev. {\bf D104}, 115006(2021).
{\it Parity violating magnetization at neutrino pair emission using trivalent
lanthanoid ions},
arXiv; 2105.1114[hep-ph].





\end{thebibliography}
\end{document}